# Mapping Nanoscale Electromagnetic Near-Field Distributions Using Optical Forces

*Fei Huang[1], Venkata Ananth Tamma[2], Zahra Mardy[1], Jonathan Burdett[2] and H. Kumar Wickramasinghe [1*]*

[1] Department of Electrical Engineering and Computer Science, 142 Engineering Tower, University of California, Irvine, USA
[2] CaSTL Center, Department of Chemistry, University of California, Irvine, USA

*Corresponding author: hkwick@uci.edu*

***Abstract***: We demonstrate the application of Atomic Force Microscopy (AFM) based optical force microscopy to map the optical near-fields with nanometer resolution, limited only by the AFM probe geometry. We map the electric field distributions of tightly focused laser beams with different polarizations and show that the experimentally measured data agrees well with the theoretical predictions from a dipole-dipole interaction model, thereby validating our approach. We further validate the proposed technique by evaluating the optical electric field scattered by a spherical nanoparticle by measuring the optical forces between the nanoparticle and gold coated AFM probe. The technique allows for wavelength independent, background free, thermal noise limited mechanical imaging of optical phenomenon with sensitivity limited by AFM performance. Optical forces due to both electric and magnetic dipole-dipole interactions can be measured using this technique.

Optical tweezers and Atomic Force Microscopy (AFM) techniques both use different types of forces at nanoscale for control and manipulation of objects. Optical tweezers, which use gradient optical forces, have been used for trapping and manipulation of atoms, nano and microstructures with important applications in physics and biology [1]-[6]. On the other hand, AFM techniques [7], which utilizes the tip-sample forces, have been used for high resolution imaging of the physical, chemical [8], magnetic [9],[10] and electrostatic [11] properties of material at the nanoscale. In addition, AFM techniques have also been used for nano-manipulation of atoms [12], nano-particles[13],[14] and biological cells [15]-[17]. Previously, we have proposed a novel microscopy technique to detect and image molecular resonances at nanometer level using just optically induced forces without requiring any far-field light collection schemes [18], [19]. This technique uses an AFM probe, such as, AFM cantilever or sharp tip mounted on tuning fork, as a sensitive force detector to measure the optically induced forces between a metallic tip and the molecule and relies on the high quality factor ($Q$) of the mechanical resonances of AFM cantilever to enhance the force signal. This novel technique extends the domain of traditional AFM based force microscopy to now include optically induced forces with potential applications in nanoscale imaging and microscopy and allows for the fundamental probing of light-matter interactions at atomic and nanoscale level directly using forces.

Previously, Near-field Scanning Optical Microscopy (NSOM) has been used to study light-matter interaction beyond the diffraction limit with important applications in many areas of nano-optics, materials science, chemistry and biology [20]. Both aperture NSOM (*a*-NSOM) [21] and aperture-less or scattering NSOM (*s*-NSOM) [22],[23] techniques have been used to map the nanoscale electric [24]-[28] and magnetic [29]-[32] field distributions. Both NSOM techniques involve the sampling of evanescent electromagnetic fields close to the sample surface due to light scattering by structured metallic probes that are brought physically close to the surface of the sample. The scattered evanescent fields are converted into propagating modes that are detected in the far-field. While many techniques have been proposed for reduction of background noise, fundamentally, NSOM techniques are limited by the use of sensitive far-field light collection and detection schemes. However, AFM based optical force microscopy technique proposed in this work directly measures the near-field as an optical force without the use of far-field detection schemes. Similar to *s*-NSOM technique [33]-[35], the optical force detection scheme is fundamentally wavelength independent and can be applied to measure both linear [18] and non-linear responses of materials [19]. The technique provides a viable and reproducible method to mechanically measure nanoscale optical phenomenon. In this work, we propose the use of AFM based Optical Force Microscopy (OFM) to investigate and map the nanoscale electromagnetic field distributions with resolution limited only by the AFM probe geometry. We experimentally map the absolute value of electric field distribution of tightly focused Gaussian beams. In addition, we map the distribution of electric field scattered by a gold nanoparticle of diameter 30 nm by measuring the optical force between the particle and a gold coated AFM probe. The measured field distributions were found to be excellent agreement with numerical calculations of focal and scattered field distributions. The experimental results were also found to be background free, limited only by the thermal noise and AFM performance.

The general scheme of an optical force microscope, detailed in Fig. 1 (a) [18], [19], can utilize AFM control using either tapping mode, shear-force feedback or tuning fork tapping mode shown Fig. 1 (e). The origin of optical forces in AFM can be well understood by considering the optical forces between the sample under measurement, modeled as a sub-wavelength magneto-dielectric particle (with electric and magnetic dipole moments $\vec{P}_p$ and $\vec{M}_p$, respectively) and the tip of the AFM probe, also modeled as a sub-wavelength magneto-dielectric particle (with electric and magnetic dipole moments $\vec{P}_t$ and $\vec{M}_t$, respectively) schematically shown in Fig. 1 (b). We assume the incident electromagnetic fields (with electric field $\vec{E}_{inc}$ and magnetic field $\vec{H}_{inc}$ which along $z$ direction) to be tightly focused at the interface between the background medium (with isotropic relative permittivity and permeability $\varepsilon_b$ and $\mu_b$, respectively) and sample (with isotropic relative permittivity and permeability $\varepsilon_s$ and $\mu_s$, respectively) shown schematically in Fig. 1 (b).

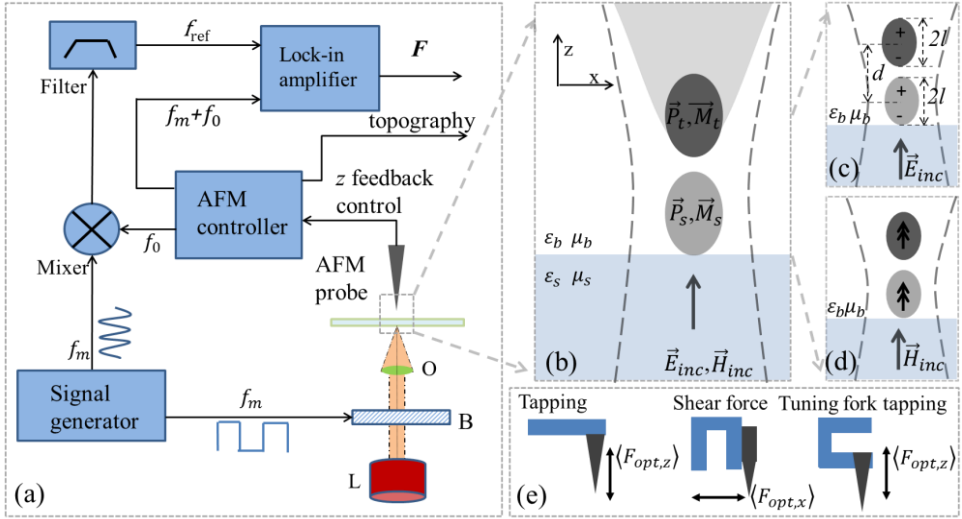

Fig. 1 (a) Schematic of Optical Force Microscopy used to measure the optically induced force **F** by locking onto the modulated optical field, generated by a laser (L). Incident light is modulated at frequency $f_m$ by a Bragg cell (B) and focused on to the sample by oil immersion objective lens (O). The cantilever or tuning fork detects this modulated field on one of its resonant frequencies while the electronic reference is generated by mixing the modulation frequency ($f_m$) and the detector self-oscillation frequency ($f_0$) in a mixer and subsequent band-pass filtering (b) Cartoon of the tip-sample interaction under illumination by tightly focused Gaussian beam with the AFM probe tip modeled as a sub-wavelength magneto-dielectric particle and the tip-sample interaction modeled as electric and magnetic dipole-dipole interaction (c) Cartoon of the tip sample electric dipole-dipole interaction (d) Cartoon of the tip sample magnetic dipole-dipole interaction (e) Three different AFM working modes include the tapping mode AFM which detects force in $z$ direction, tuning fork AFM working in shear force mode detects force in $x$ direction while tuning fork AFM working in tapping mode detects force in $z$ direction.

The total force experienced by the AFM probe tip is $\vec{F}_{tot} = \vec{F}_{int} + \vec{F}_{opt}$, where, $\vec{F}_{int}$ is the total tip-sample interaction forces consisting of all van der Waals forces, meniscus forces, chemical and Casimir forces, and $\langle\vec{F}_{opt}\rangle$ is the time-averaged optical force on the AFM probe tip due to its interaction with the incident field and particle dipole. Due to the presence of both electric and magnetic dipoles, $\langle\vec{F}_{opt}\rangle$ can be written as the sum, $\langle\vec{F}_{opt}\rangle = \langle\vec{F}_e\rangle + \langle\vec{F}_m\rangle + \langle\vec{F}_{e-m}\rangle$ [36]-[38], where, $\langle\vec{F}_e\rangle = \frac{1}{2}\Re\{\vec{P}_t(\nabla \cdot \vec{E}_{t,loc})\}$ is the time-averaged force experienced due to the electric dipole with spatially non-varying moment $\vec{P}_t$, $\langle\vec{F}_m\rangle = \frac{1}{2}\Re\{\vec{M}_t(\nabla \cdot \vec{B}_{t,loc},)\}$ is the time-averaged force experienced due magnetic dipole with spatially non-varying moment $\vec{M}_t$, $\langle\vec{F}_{e-m}\rangle = -\frac{1}{2}\Re\left\{\frac{2k^4}{3}\sqrt{\frac{\mu_b}{\varepsilon_b}}(\vec{P}_t \times \vec{M}_t)\right\}$ is the interaction force due to coupling between the electric and magnetic dipoles [36]-[38] and *k* is the wavenumber of the incident wave. Here, $\vec{E}_{t,loc}$ and $\vec{B}_{t,loc}$ are the local electric and magnetic fields experienced by the AFM probe tip respectively and are given by the sum of the incident field and the fields scattered by the particle.

Proper choice of $\vec{P}_t$ and $\vec{M}_t$ would allow for the detection of optical electric and/or magnetic fields using forces. For example, the use of a metal coated tip supporting large values of $\vec{P}_t$ could be used to measure the electric dipole-dipole interaction forces schematically shown in Fig. 1 (c). Due to the weak nature of magnetism in materials at optical frequencies, magnetic dipole-dipole interaction forces could be detected by measuring the interaction forces between nanostructures supporting artificial magnetism. In particular, it is well known that nano-spheres composed of dense dielectric materials support the magnetic Mie resonance. In addition, metal nanoparticle clusters with specific geometric arrangements have been used to demonstrate magnetic activity at optical frequencies. Therefore, measurement of forces between two magnetic dipoles oscillating at optical frequencies, shown schematically in Fig. 1 (d), could be achieved by measuring the interaction forces between two nanostructures supporting artificial magnetism at the chosen frequency. For example, optical forces between two magnetic dipoles could be measured by measuring the forces between two interacting silicon nano-spheres with one of the spheres placed on a substrate and the other attached to the end of an AFM probe. Experimental work is currently underway to experimentally measure the optical forces between two silicon nano-spheres and also map the spatial distributions incident focal magnetic fields.

We chose to map the nanoscale electric field distributions due to tightly focused Gaussian beams by using a gold coated AFM probe and measuring the forces between the induced dipole at the apex and its image dipole in glass as substrate. Assuming zero magnetic dipole moment, both $\langle \vec{F}_m \rangle$ and $\langle \vec{F}_{e-m} \rangle$ are neglected. The experiments were performed using a tapping force AFM and therefore only $\langle F_{opt,z} \rangle$ is the component of the total force that is of interest since that would be the only component detected by the AFM cantilever and can be calculate as [46]

$$< F_{opt,z} >= \frac{24\pi\varepsilon_0 d\alpha'_{tip,x}\alpha'_{img,x}}{(d^2+l'^2)^{5/2}}|E_{inc,x}|^2 + \frac{24\pi\varepsilon_0 d\alpha'_{tip,y}\alpha'_{img,y}}{(d^2+l'^2)^{5/2}}|E_{inc,y}|^2 + \frac{8\pi\varepsilon_0\alpha'_{tip,z}\alpha'_{img,z}(3d^2+l^2)}{(d^2-l^2)^3}|E_{inc,z}|^2 \quad (3)$$

where, *2l* is the length of the major axis (along the *z*-axis) and *2l'* is the length of the minor axes, $\alpha'_{tip,x}$, $\alpha'_{tip,y}$ and $\alpha'_{tip,z}$ are the real part of electric polarizabilities of the prolate spheroid along the *x*, *y* and *z* axis [40] and $\alpha'_{img,x}$, $\alpha'_{img,y}$ and $\alpha'_{img,z}$ are the real part of electric polarizabilities of the image dipole along the *x*, *y* and *z* axis. We note that the image dipole polarizabilites are proportional to the polarizabilities of the tip by the scaling factor $(\varepsilon_{sub} - \varepsilon_{air})/(\varepsilon_{sub} + \varepsilon_{air})$ [20]. However, as shown schematically in Fig. 1 (e), shear force AFM feedback mode could be used to detect the *x*-component $< F_{opt,x} >$ of the total force. From (3), we note that the image force is a sum of three separate contributions from the interaction of the *i*[th]-component of the field with the tip, where *i* = *x, y, z* and this property could be used to decompose the polarization dependence of the optical field under measurement. The electric field distribution at the focus of a tightly focused laser beam was calculated using the plane wave expansion method [20]. We assumed $\vec{E}_{inc}$, which is the electric field strength of light incident on the high NA objective lens, to be polarized along the *x*-axis. Since the focal field distribution for $\vec{E}_{inc}$ is known to contain components along both the *x* and *z* axes [20], [39], $< F_{opt,z} >$ contains force terms due to both *x* and *z* oriented electric dipoles.

The experimental setup was built around a commercial AFM (Veeco Caliber) operating in the tapping mode and the schematic of the experimental setup is shown in Fig. 1 (a). The optical field under measurement, generated by a laser of suitable wavelength, was modulated using a Bragg cell modulator at the modulation frequency $f_m$ and focused on the sample by using an oil immersion objective with NA = 1.45. The incident optical field induces a dipole at the end of the AFM probe shown in the schematic of the tip-sample interaction under illumination by a tightly focused laser beam and is modulated at frequency $f_m$. Thus the attractive gradient optical force (***F***) between the induced dipole on the tip and its image dipole on the substrate was also modulated at frequency $f_m$ which in turn modulated the AFM cantilever first mechanical resonance frequency $f_0$ generating sidebands at $f_0+f_m$ and $f_0-f_m$. We chose to locate the sideband at $f_0+f_m$ on top of the second mechanical resonance frequency of the AFM cantilever thereby utilizing the high $Q$ of the second resonance of the AFM cantilever [41] to enhance optical force signal. The optical image force signal was then detected by use of a lock-in amplifier while the electronic reference was obtained by mixing the modulating frequency $f_m$ and the cantilever resonance frequency $f_0$ in a balanced mixer followed by a bandpass filter centered at $f_0+f_m$. In our experiments, we chose $f_m$ = 360 kHz and the sideband frequency $f_0+f_m \sim$ 425 kHz. We experimentally mapped the focal field distribution at 640 nm using an oil immersion objective with NA = 1.45 on a 0.16 mm thick cleaned glass microscope cover slide. The input light was polarized along $x$-axis using a polarizer. The objective back aperture was completely over filled to ensure a tight focal spot. The tapping amplitude of the AFM cantilever was set to 40 nm and the focal spot was raster scanned to simultaneously obtain spatial distributions of topography and optical image force plotted in Fig. 2 (a) and (b) respectively. The topography shows a surface roughness (rms) of 0.56 nm for the cleaned glass slide. The optical image force distribution exhibits two distinct spots and the experimentally measured focal spot size of 550 nm agrees well with the theoretical prediction of 539 nm. To compare, the image force was numerically evaluated at 640 nm using (7) and the results plotted in Fig. 2 (c) agreed well with the experimentally measured data. Further, a line trace extracted from Fig. 2 (b) (along line a-a') was compared with the trace extracted from Fig. 3 (c) (along line b-b') in Fig. 2 (d) and the experimentally measured image force data agreed well with the numerical calculations. The numerical data was fit to the experimentally measured data using the following parameters: $l$ = 15 nm, $l'$ =3.5 nm, $d = 2l$ nm (assume tip contact with glass surface) and $\varepsilon_s = 1.47$ .

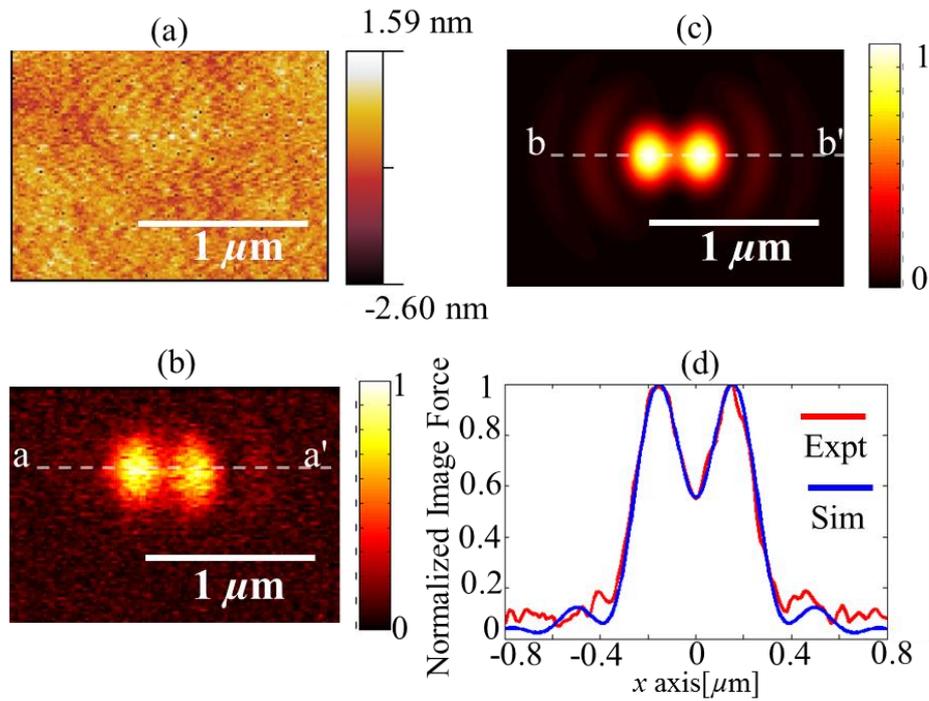

Fig. 3. Spatial distributions of (a) Topography and (b) Normalized optical image force measured experimentally on a cleaned glass microscope cover slide at 640 nm with light polarized along $x$-axis (c) Numerical calculations of normalized image force obtained by evaluating (3) (d) Comparison of image force trace obtained experimentally (line a-a' in (b)) and simulations (line b-b' in (c)) showing very good agreement.

To further verify our experimental findings, we experimentally mapped the focal field distribution at 660 nm using an oil immersion objective with NA = 1.45 on a 0.16 mm thick cleaned glass microscope cover slide. Spatial distributions of normalized optical image force plotted in Fig. 3 (a) to (d) were obtained for light polarized along $x$ axis, $y$ axis (rotated using a half wave plate), azimuthal polarization and radial polarization (obtained using Arcoptix radial polarizer), respectively. The field distributions for the azimuthal and radial polarizations plotted in Figs. 3 (c), (d) agree well with the theoretical predictions [42]. From the inset in Fig. 3 (c), the full width half maximum (FWHM) of the null in the field distribution is 337 nm and agrees well with previously published results [42].From the inset in Fig. 3 (d) for radially polarized light, the full width half maximum (FWHM) of the focal spot is 356 nm and agrees well with previously published results [42]. We note that the focal field distribution of tightly focused Gaussian beam with radial polarization consists of two distinct regions with orthogonal polarizations. The central spot is purely longitudinally polarized ($E_z$) field and the circular ring surrounding the central ring is purely transverse polarized ($E_\rho$) [42].

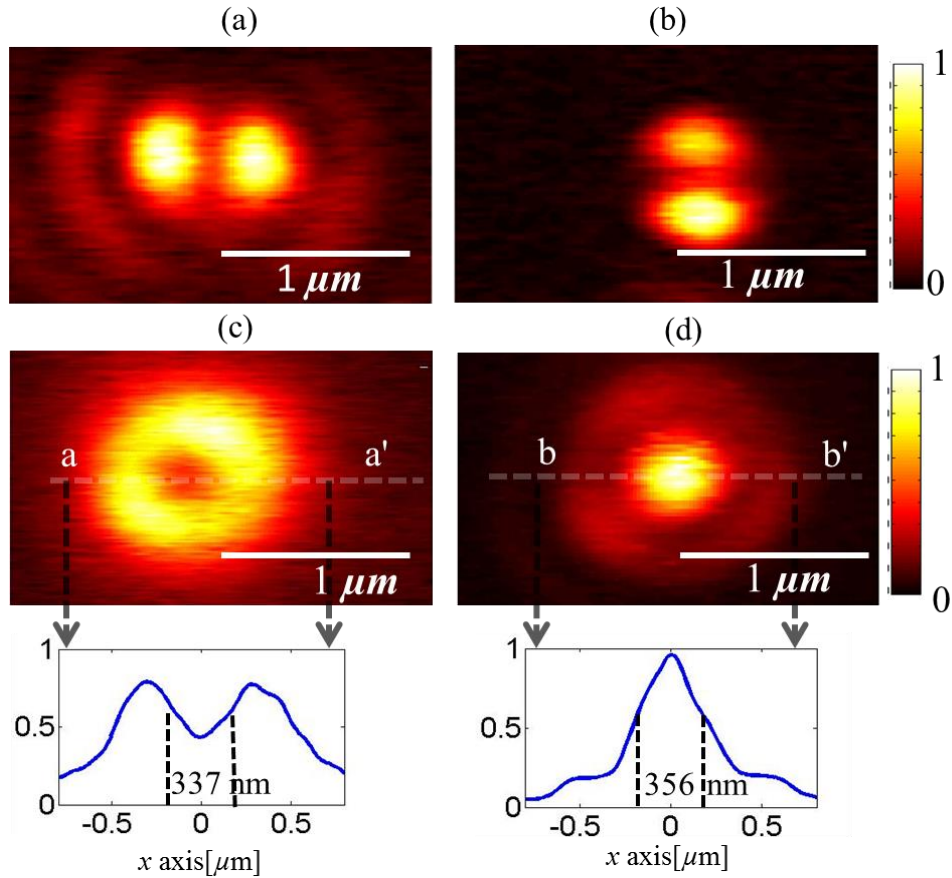

Fig. 3. Spatial distributions of (a) normalized optical force with polarization along *x* axis (b) normalized optical image force with polarization rotated in-plane by 90° when compared to (a) (c) normalized optical force with azimuthal polarized light (d) normalized optical force with radially polarized light.

We performed further experiments to further understand the contrast mechanism in the optical image force results plotted in Fig. 3, 4. We experimentally map the focal field distributions of *x*-polarized input light at 685 nm using an oil immersion objective with NA = 1.45 on a 0.16 mm thick cleaned glass microscope cover slide for two different amplitude setpoints while maintaining the same AFM free tapping amplitude of 54 nm. This ensured that the separation between the tip and sample (*d*-2*l*) could be accurately controlled with nanometer precision. Spatial distributions of the normalized optical force for two different amplitude setpoints of 52.06 nm and 52.40 nm are plotted in Figs. 4 (a) and (b), respectively. The spatial distribution of normalized optical force for the setpoint of 52.06 nm compares well with the distributions plotted in Figs. 2 and 3 whereas we do not observe any optical force signal for the measurement scans performed with a setpoint of 52.40 nm. The scans in Figs. 4(a) and (b) were performed using the same gold coated AFM probe and thereafter, an AFM amplitude-distance experiment was performed whose results are plotted in Fig. 4 (c) and (d), respectively. We note that after performing the AFM amplitude-distance curve experiment, the gold coated probe could no longer be used to obtain the spatial optical force distributions as plotted in Fig. 2(a) indicating severe damage to the apex of the gold coated probe. From the linear fit to the AFM amplitude-distance curve plotted in Fig. 4 (d), we estimate that varying the amplitude setpoint from 52.06 nm to 52.40 nm resulted in a tip-sample separation (*d*-2*l*) of 1.5 ± 0.2 nm. Indeed, such a small tip-sample separation was found to be sufficient for a dramatic decrease in the optical force signal as evident from the plots in Figs. 4 (a) and (b). However, the estimated tip-sample separation of 1.5 ± 0.2 nm is very small compared to the confocal length of the tightly focused optical spot [20] and theoretically, we expect the optical force to have very similar spatial distributions for such small changes in tip-sample distance. The distance dependence of optical image force plotted in Fig. 4 is a direct verification of dipole-dipole interaction theory.

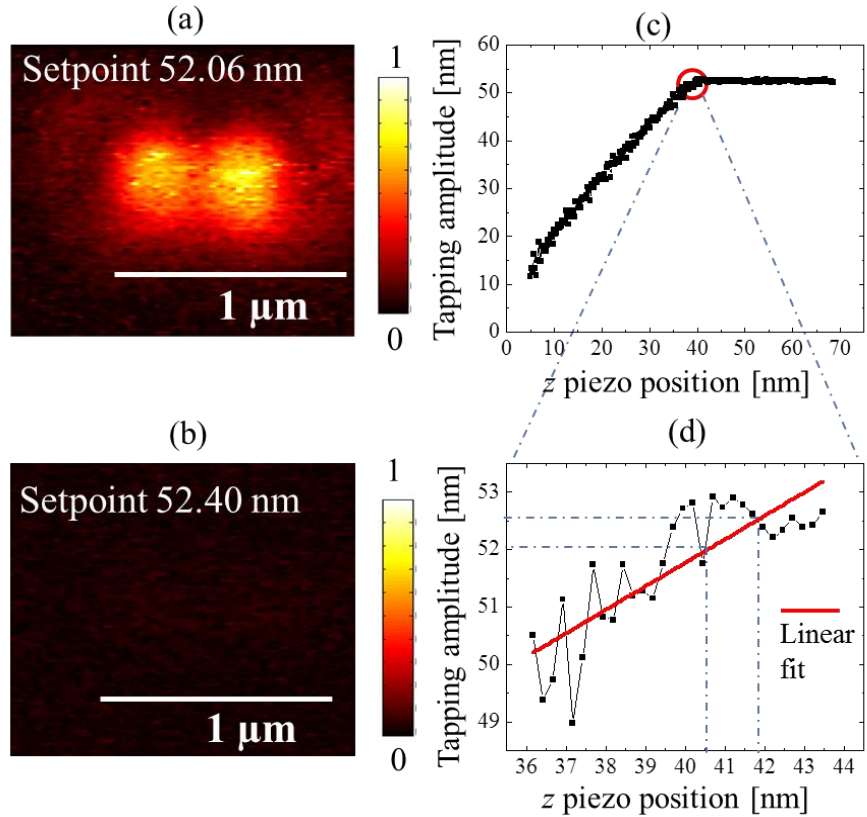

Fig.4. Spatial distribution of optical image force on cleaned glass at 685 nm with light polarized along *x*-axis measured at an amplitude setpoint of (a) 52.06 nm and (b) 52.40 nm (c) AFM amplitude-distance curve obtained immediately after performing both measurements using same AFM probe (d) Plot of zoomed-in AFM amplitude-distance curve in (c) with a linear fit.

Finally, we performed experiments to map the spatial distribution of electric field scattered by a single gold nanoparticle of diameter 30 nm (Sigma-Aldrich) which was dropcast onto a cleaned glass surface. The nanoparticle was placed in the center of the focal spot obtained using an oil immersion objective with NA = 1.45, with radially polarized input light of wavelength 660 nm. The nanoparticle was then raster scanned using a gold coated AFM probe to obtain spatial distributions of topography and normalized optical image force plotted in Fig.5 (a) and (b), respectively. In Fig. 5 (b), we present the optical force maps of two 30 nm gold nanoparticles illuminated by incident focused light with two different and orthogonal polarizations of light. First, a single nanoparticle was positioned to be within the central spot with purely longitudinal fields ($E_z$) and the resulting field distribution shows typical plasmonic hotspot on the nanoparticle. Another nanoparticle was found to be located within the circular ring with purely transverse fields ($E_\rho$) and the resulting field distribution shows a typical dipole like pattern on the nanoparticle. Line traces of the topography and normalized optical forces were extracted from Fig. 5 (a) (along line a-a') and Fig. 5 (b) (along line b-b' and d-d') and plotted in Figs. 5 (c), 5 (d) and 5 (f), respectively. The feature width in the line trace of topology plotted in Fig. 5 (c) is 94 nm and the broadening was attributed to the topographic convolution with probe apex geometry [43]. The width of the measured optical force due to the nanoparticle located in the longitudinal ($E_z$) field and extracted from the normalized optical force plotted in Fig. 5 (d) was found to be considerably smaller at 71 nm, or ~$\lambda/9.29$ at a wavelength of 660 nm. To further study the observed results and obtain a qualitative understanding of the physics, we performed numerical calculations using the commercial finite element solver COMSOL. The AFM probe tip was modeled as a gold nanoparticle of diameter 5 nm and followed the theoretical trace of the AFM topography over the 30 nm diameter gold particle plotted in Fig 5 (e). The force experienced by the 5 nm gold nanoparticle as it traversed the theoretical curve around the 30 nm gold nanoparticle was calculated by evaluating the Maxwell stress tensor using the electric fields obtained from COMSOL simulations. Line traces of calculated forces for both the longitudinal ($E_z$) and transverse ($E_\rho$) fields using the same theoretical topology trace are also plotted in Figs. 5 (d) and 5 (f), respectively along with experimental data. The results of the simple quantitative simulations are in good agreement with experimental data and both similar trends. The experimental data was found to be slightly broadened and can be explained by considering the conical shape of the AFM probe. The dips in the experimental force curves on both sides of the particles, seen in both Figs. 5 (d) and 5 (f), can be result due to the net force exerted on the 5 nm gold nanoparticle as it interacts with both the 30 nm gold nanoparticle to its side and with the incident electric field to its bottom.

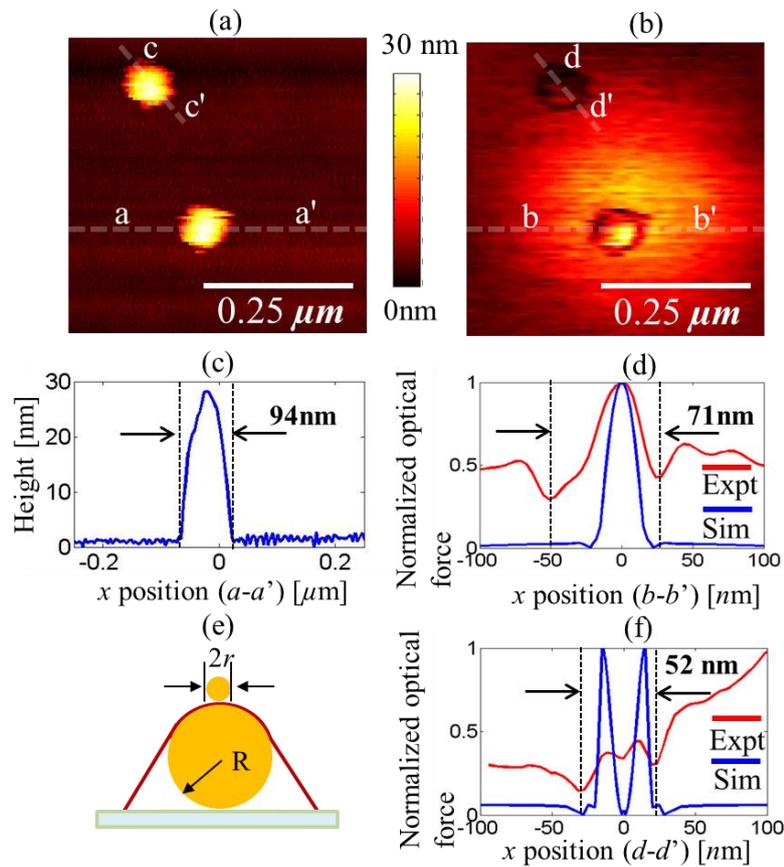

Fig. 5. Spatial distributions of (a) topography and (b) normalized optical image force measured experimentally on a gold nanoparticle of diameter 30 nm on a cleaned glass surface at 660 nm with radially polarized input light (c) line scan of topography indicating feature size of 94 nm (d) line scan of normalized optical force with 30 nm gold nanoparticle in the $E_z$ incident field indicating feature size of 71nm and compared with the line scan from numerical calculations (e) Model used in numerical calculations with gold nanoparticle of diameter $2r$=5nm scanning over the gold nanoparticle of diameter 30 nm and following the theoretical trace of the AFM topography (f) line scan of normalized optical force with 30 nm gold nanoparticle in the transverse incident field and compared with the line scan from numerical calculations.

In conclusion, we experimentally demonstrated the application of optical force and AFM based technique to map the optical near-field distributions. The spatial resolution is limited only by the AFM probe geometry. We have mapped the spatial electric field distributions of tightly focused laser beams with linear, radial and azimuthal polarizations. We have derived a dipole-dipole theory to model the image force and show that the experimentally measured data agrees well with the theoretical predictions, thereby validating our approach. In addition, we have mapped the spatial electric field distribution of a gold nanoparticle for input radial polarization and report a sub-wavelength electric field hot spot measured using optical forces.

**Acknowledgement:** The authors thank Yinglei Tao, Janhoon Jahng, Dimitry A.Fishman, Jordan Brocious, Xaeowei Li, Nicholas Sharac, Eric O.Potma for helpful discussions. The experimental and theoretical collaboration is supported by the NSF Center for Chemical Innovation, Chemistry at the Space-Time Limit (CaSTL) under Grant No. CHE-0802913.

**Contributions:** FH conducted experiments with assistance from VAT and JB who helped with experimental setup. FH and VAT analyzed experimental data. VAT and ZM performed COMSOL simulations. HKW conceived and supervised the project. FH, VAT and HKW wrote the manuscript.

**References**

1. Ashkin, A. Acceleration and trapping of particles by radiation pressure. *Phys. Rev. Lett*. **24**, 156–159(1970)

2. Ashkin, A. Optical levitation by radiation pressure. *Appl.Phys. Lett.* **19**, 283–285(1971).

3. Ashkin, A. ,Dziedzic, J.M.,Bjorkholm,J.E.& Chu,S. Observation of a single-beam gradient force optical trap for dielectric particles. *Opt. Lett*. **11**, 288–290 (1986).

4. Ashkin,A., Dziedzic,J.M.&Yamane, T. Optical trapping and manipulation of single cells using infrared laser beams. *Nature* **330**, 769–771(1987).

5. Grier, D. G. A revolution in optical manipulation. *Nature* **424**, 810-816 (2003).

6. Moffitt, J. R., Chemla, Y. R., Smith, S. B. & Bustamante, C. Recent Advances in Optical Tweezers. *Annu.Rev. Biochem*. **77**,205–28(2008).

7. Martin, Y. , Williams, C. C.& Wickramasinghe, H. K. Atomic force microscope-force mapping and profiling on a sub 100Å scale. *J. Appl. Phys.* **61**, 4723 (1987).

8. Gross, L. et al. Bond-Order Discrimination by Atomic Force Microscopy. *Science* **337**, 1326–1329 (2012).

9. Martin Y. & Wickramasinghe H.K. Magnetic Imaging by Force Microscopy with 1000A Resolution. *Appl. Phys. Lett.* **50** (20), 1455–1457(1987).

10. Rugar, H. et al. Magnetic force microscopy: General principles and application to longitudinal recording media. *J. Appl. Phys.***68**, 1169 (1990).

11. Nonnenmacher, M. , O'Boyle, M. P. & Wickramasinghe H. K. Kelvin probe force microscopy. *Appl. Phys. Lett*. **58** (25), 2921. (1991).

12. Bamidele, J.et al. Chemical tip fingerprinting in scanning probe microscopy of an oxidized Cu(110) surface. *Phys. Rev. B* **86**, 155422 (2012).

13. Kim, S. ,Shafiei, F. , Ratchford , D. & Li, X. Controlled AFM manipulation of small nanoparticles and assembly of hybrid nanostructures. *Nanotechnology* **22**, 115301(2011).

14. Sharma , S. et al. Structural-Mechanical Characterization of Nanoparticle Exosomes in Human Saliva, Using Correlative AFM, FESEM, and Force Spectroscopy. *ACS Nano* **4 (4)**,1921–1926 (2010).

15. Ando, T. , Uchihashi , T.&Fukuma, T. High-speed atomic force microscopy for nano-visualization of dynamic biomolecular processes. *Prog.Surf. Sci*. **83**, 337–437 (2008).

16. Nawarathna, D. , Unal, K. & Wickramasinghe, H. K. Localized electroporation and molecular delivery into single living cells by atomic force microscopy. *Appl. Phys. Lett.***93**, 153111 (2008).

17. Nawarathna, D. , Turan, T. & Wickramasinghe, H. K. Selective probing of mRNA expression levels within a living cell. *Appl. Phys. Lett.* **95**, 083117 (2009).

18. Rajapaksa, I. , Uenal, K. & Wickramasinghe, H. K. Image force microscopy of molecular resonance: A microscope principle. *Appl. Phys. Lett*. **97**, 073121 (2010).

19. Rajapaksa, I. & Wickramasinghe, H. K. Raman spectroscopy and microscopy based on mechanical force detection. *Appl. Phys. Lett*. **99**, 161103 (2011).

20. Novotny, L. & Hecht, B. *Principles of Nano-Optics*. (Cambridge Univ. Press, Cambridge, UK, 2006).

21. Hecht, B. et al. Scanning near-field optical microscopy with aperture probes: Fundamentals and applications. *J. Chem. Phys*. **112**, 7761 (2000).

22. Zenhausern, F. , Martin, Y. & Wickramasinghe, H. K. Scanning Interferometric Apertureless Microscopy: Optical Imaging at 10 Angstrom Resolution. *Science* **269**, 1083 (1995).

23. Schnell, M. et al. Controlling the Near-Field Oscillations of Loaded Plasmonic Nanoantennas. *Nat. Photonics* **3**, 287−291 (2009).

24. Zhang, Z. , Ahn, P. , Dong, B. , Balogun, O. & Sun, C. Quantitative imaging of rapidly decaying evanescent fields using plasmonic near-field scanning optical microscopy. *Scientific Report*, **Vol. 3**, 2803(2013).

25. Veerman, J.A., Garcia-Parajo, M.F. , Kuipers, L. & van Hulst. N.F. Single molecule mapping of the optical field distribution of probes for near-field microscopy. *J.Microsc*.**194**,477-482(1999).

26. Bauer, T. ,Orlov, S. , Peschel, U. , Banzer, P. & Leuchs, G. Nanointerferometric amplitude and phase reconstruction of tightly focused vector beams. *Nature Photon.* **8**, 23–27 (2014).

27. Kohlgraf-Owens, D. C, Greusard, L. , Sukhov, S. ,Wilde Y. De.& Dogariu A. Multifrequency nearfield scanning optical microscopy. *Nanotechnology* **25** 035203 (2014).

28. Fleischer, M. et al. Three-dimensional optical antennas: Nanocones in an apertureless scanning near-field microscope. *Appl. Phys. Lett*. **93**, 111114 (2008).

29. Feber, B. le, Rotenberg, N. , Beggs D. M. & Kuipers, L. Simultaneous measurement of nanoscale electric and magnetic optical fields. *Nature Photon.* **8**, 43–46 (2014).

30. Kihm, H.W. et al. Bethe-hole polarization analyser for the magnetic vector of light. *Nat. Commun.* **2**, 451(2011).

31. Burresi, M. et al. D. Probing the magnetic field of light at optical frequencies. *Science* **326**,5952, 550-553 (2009).

32. Kihm, H. W. et al.Optical magnetic field mapping using a subwavelength aperture. *Opt. Express* **21**, 5625-5633 (2013).

33. Huth, F. et al. Nano-FTIR Absorption Spectroscopy of Molecular Fingerprints at 20 nm Spatial Resolution. *Nano Lett*. **12**, 3973–3978(2012).

34. Pettinger, B., Schambach, P., Villagomez, C. J. & Scott, N.Tip-Enhanced Raman Spectroscopy: Near-Fields Acting on a Few Molecules. *Annu. Rev. Phys. Chem*. **63**, 379–99(2012).

35. Chen, C., Hayazawa, N., Kawata, S. A 1.7 nm resolution chemical analysis of carbon nanotubes by tip-enhanced Raman imaging in the ambient. *Nat.Commun.* **5**, 3312(2012).

36. Nieto-Vesperinas, M. , Sáenz, J. J., Gómez-Medina, R. & Chantada, L. Optical forces on small magnetodielectric particles. *Opt. Express* **18**, 11428-11443 (2010).

37. Jay, K., Chaumet, P.C., Langtry, T. N.& Rahmani, A. Optical binding of electrically small magnetodielectric particles. *J. Nanophoton*. **4(1)**, 041583 (2010).

38. Chaumet, P. C. & Rahmani, A. Electromagnetic force and torque on magnetic and negative-index scatterers. *Opt. Express* **17**, 2224–2234 (2009).

39. Bohren, C. F., Huffman, D. R. Absorption and Scattering of Light by Small Particles. (*Wiley Interscience*, New York, 1983)

40. K. Furusawa, N. Hayazawa, T. Okamoto, T. Tanaka, S. Kawata, Generation of Broadband Longitudinal Fields for Applications to Ultrafast Tip-Enhanced Near-Field Microscopy. *Opt. Express* **19**, 25328−25336 (2011)

41. Garcia, R. & Herruzo, E. T. The emergence of multifrequency force microscopy. *Nature Nanotech.* **7**, 217–226 (2012).

42. Youngworth, K. S. & Brown, T. G. Focusing of high numerical aperture cylindrical-vector beams. *Opt. Express* **7**, 77-87 (2000).

43. Chasiotis, I. Atomic Force Microscopy in Solid Mechanics. Springer Handbook of Experimental Solid Mechanics, 409-444 (2008).

44. Hallock, A. J., Redmond, P. L. & Brus, L. E. Optical forces between metallic particles. *Proc. Nat. Acad. Sci. U.S.A.* **102**, 1280–1284 (2004).

45. Johnson, P.B. & Christy, R.W. Optical Constants of the Noble Metals. *Phys. Rev. B* **6**, 4370−4379 (1972).

46. Supplementary material with calculations for methods, electric dipole-dipole interaction force, magnetic dipole-dipole interaction force and estimation of measured force and force gradients.

**Supplementary notes:**

I. Calculation of electric dipole-dipole interaction force
II. Calculation for magnetic force
III. Estimate of the detected force and force gradient
IV. Methods

I. <u>Calculation of electric dipole-dipole interaction force</u>

Since we chose to map the nanoscale electric field distributions, the force due to magnetic dipole-dipole interaction is neglected. The total time average electric force [1] is

$$\langle \vec{F}_e \rangle = \frac{1}{2} \Re \{ \vec{P}_{t,loc} (\nabla \cdot \vec{E}_{t,loc}) \} \qquad (1)$$

Where, $\vec{E}_{t,loc}$ is the total electric field experienced by the AFM probe tip and is given by the sum of the incident field and the fields scattered by the particle dipole [2]. Equation (1) can also be written as

$$\langle \vec{F}_e \rangle = \frac{1}{4} \alpha' \nabla |E_{t,loc}|^2 + \frac{k}{2n} \alpha'' \Re \left( \vec{E}_{t,loc} \times \vec{B}^*_{t,loc} \right) + \frac{1}{2} \alpha'' \Im [(E^*_{t,loc} \cdot \nabla) E_{t,loc}] \qquad (2)$$

where, the electric polarizability of the magneto-dielectric particle $\alpha = \alpha' + i\alpha''$ with real and imaginary parts, $\alpha'$ and $\alpha''$, respectively. In (2), we associate the first term with the electric dipole-dipole interaction force, the second term with the scattering force, and the third term with a curl force associated to the non-uniform distribution of the spin density of the electric field. We note that since the scattering force along the *z*-direction was constant in the vicinity of the focal spot and considering the typical vibration of the AFM cantilever was 40 nm, the scattering force was not detected. Also, the curl force was zero due to the uniform distribution of the spin density of the electric field in the focal plane.

Experiments were performed using a tapping force AFM and therefore only $\langle F_{opt,z} \rangle$ is the component of the total force that is of interest since that would be the only component detected by the AFM cantilever and is expanded in terms of the fields as

$$\langle F_{opt,z} \rangle = p_{tip,x} \frac{\partial E_{tip,z}}{\partial x} + p_{tip,y} \frac{\partial E_{tip,z}}{\partial y} + p_{tip,z} \frac{\partial E_{tip,z}}{\partial z} \tag{3}$$

where, $p_{tip,x}$, $p_{tip,y}$ and $p_{tip,z}$ the components of the electric dipole moment along the *x*, *y* and *z* axis respectively and $E_{tip,z}$ is *z* components of the total electric field on the AFM probe tip. We modeled the interacting region of the tip as a prolate spheroid with $l > l'$, where, *2l* is the length of the major axis (along the *z*-axis) and *2l'* is the length of the minor axes [3]. We note that $p_{tip,x} = p_{tip,y}$ assuming symmetry of the AFM tip. The configurations of the dipole-dipole interaction model for $E_z$ and $E_x$ incident fields of tip dipole and image dipole are shown in Figs. S1 (a) and (b) respectively.

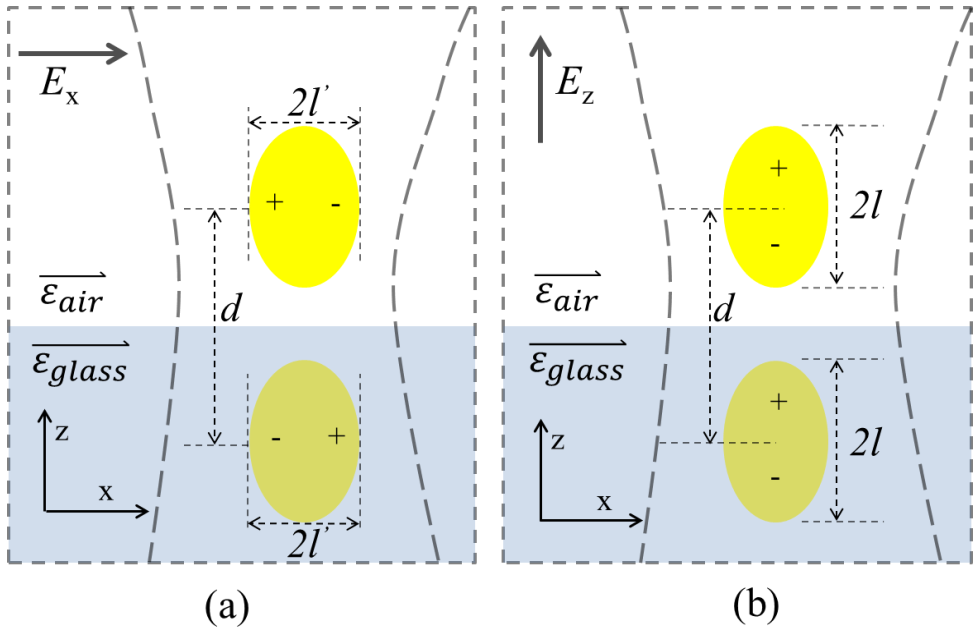

Fig. S1: Tip sample interaction due to electric field modeled as the dipole-dipole interaction along (a) *x* direction, (b) *z* direction.

Assuming $l, l' \ll \lambda$, using the electrostatic approximation [2], we obtain

$$\Re\{p_{tip,x}\}\frac{\partial E_z}{\partial x} \approx \frac{24\pi\varepsilon_0 d \alpha'_{tip,x} \alpha'_{img,x}}{(d^2+l'^2)^{5/2}} |E_{inc,x}|^2 \tag{4}$$

$$\Re\{p_{tip,y}\}\frac{\partial E_z}{\partial y} \approx \frac{24\pi\varepsilon_0 d \alpha'_{tip,y} \alpha'_{img,y}}{(d^2+l'^2)^{5/2}} |E_{inc,y}|^2 \tag{5}$$

$$\Re\{p_{tip,z}\}\frac{\partial E_z}{\partial z} \approx \frac{8\pi\varepsilon_0 \alpha'_{tip,z} \alpha'_{img,z}(3d^2+l^2)}{(d^2-l^2)^3} |E_{inc,z}|^2 \tag{6}$$

where, $\alpha'_{tip,x}$, $\alpha'_{tip,y}$ and $\alpha'_{tip,z}$ are the real part of electric polarizabilities of the prolate spheroid along the *x*, *y* and *z* axis [3] and $\alpha'_{img,x}$, $\alpha'_{img,y}$ and $\alpha'_{tip,z}$ are the real part of electric polarizabilities of the image dipole along the *x*, *y* and *z* axis. We note that the image dipole polarizabilites are proportional to the polarizabilities of the tip by the scaling factor $(\varepsilon_{sub} - \varepsilon_{air})/(\varepsilon_{sub} + \varepsilon_{air})$ [2]. In deriving (4)-(6), we have neglected terms with fast spatial decay rates of $d^{-7}$ as our experimental setup is unable to detect signals with such rapid decay. Finally, the total detectable optical image force experienced by the tip is given by the sum of (4)-(6)

$$<F_{opt,z}> = \frac{24\pi\varepsilon_0 d\alpha'_{tip,x}\alpha'_{img,x}}{(d^2+l'^2)^{5/2}}|E_{inc,x}|^2 + \frac{24\pi\varepsilon_0 d\alpha'_{tip,y}\alpha'_{img,y}}{(d^2+l'^2)^{5/2}}|E_{inc,y}|^2 + \frac{8\pi\varepsilon_0 \alpha'_{tip,z}\alpha'_{img,z}(3d^2+l^2)}{(d^2-l^2)^3}|E_{inc,z}|^2 \quad (7)$$

## II. Calculation for magnetic force

Assuming the interaction region of the tip has magnetic dipole moment $\vec{M}_t$, and the sample particle has magnetic dipole moment $\vec{M}_p$. The optical magnetic interaction force between the tip and sample particle, with the incident magnetic field $\vec{H}_{inc}$ can be written as [1]

$$\langle \vec{F}_m \rangle = \frac{1}{2}\Re\{\vec{M}_{t,loc}(\nabla \cdot \vec{B}_{t,loc})\} \quad (9)$$

where, $\vec{B}_{t,loc}$ is the total magnetic field experienced by the AFM probe tip and is given by the sum of the incident magnetic field and the magnetic fields scattered by the particle dipole. The force due to magnetic dipole-dipole interaction is expanded as [1]

$$\langle \vec{F}_m \rangle = \frac{1}{4}\alpha'_m \nabla |B_{t,loc}|^2 + \frac{k}{2}n\alpha''_m \Re(\vec{E}_{t,loc} \times \vec{B}^*_{t,loc}) + \frac{1}{2}\alpha''_m \Im[(B^*_{t,loc} \cdot \nabla)B_{t,loc}] \quad (10)$$

where, the magnetic polarizability of the magneto-dielectric particleis $\alpha_m = \alpha'_m + i\alpha''_m$ with real and imaginary parts, $\alpha'_m$ and $\alpha''_m$, respectively. In (10), we associate the first term with the magnetic dipole-dipole interaction force, the second term with the scattering force, and the third term as a curl force associated with the non-uniform distribution of the spin density of the magnetic field. We note that since the scattering force along the z-direction was constant considering the extremely small vibration of the AFM cantilever (~ 40 nm) when compared with the confocal length of the focused field, the scattering force was not detected. Also, the curl force was zero due to the uniform distribution of the spin density of the magnetic field. Using electromagnetic duality, we can simply perform the substitution [2][E, H, $\mu_0\mu, \varepsilon_0\varepsilon, p$] $\rightarrow$ [H, $-$E, $\varepsilon_0\varepsilon, \mu_0\mu, \mu M$] in (7) and obtain the magnetic dipole-dipole interaction force.

## III. Estimate of the detected force and force gradient

The thermal noise amplitude $N$ of the cantilever at the second resonance [4] is given by $N^2 = 4KTBQ_2/\omega k_2$, where, $k_2$ is the cantilever stiffness at the second resonance, $K$ is Boltzmann constant, $B$ is the system bandwidth, $Q_2$ is the Q of the cantilever second resonance, $T$ is the absolute temperature and $\omega$ is the resonance frequency. For $T = 300K, B = 10Hz, Q_2 = 200, k_2 = 39.31 * k_1 = 62.898\ N/m$ [5] and $\omega = 2\pi * 425\ KHz$, we obtain $N = 4.441 \times 10^{-4}$ nm, which gives the minimum detectable second resonance amplitude. Using $F = k_2 x/Q_2$, the corresponding force for the noise second resonance is $F = 1.39 \times 10^{-13}$ N. Force gradient can be calculated using $F' = \Delta A/A * 2k_1/Q_1$. $\Delta A/A$ is the measured modulation depth, For all the data presented in the manuscript, we measured the modulation depth to be -70 db (or signal to carrier ratio $3.162 \times 10^{-4}$), $Q_1 = 100$, $k_1 = 1.6\ N/m, F' = 2.023 \times 10^{-5}$.

## V. Methods

Gold coated AFM cantilever probes were prepared by sputter coating commercial bare Silicon AFM cantilever probes (AppNano Forta $k$ = 1.6 N/m, $f_0$ ~ 65 kHz) with 25 nm gold on a 2 nm chromium adhesion layer. Cleaned glass slides were prepared by rinsing 1.6 mm thick glass cover slides in acetone, methanol and isopropanol consecutively. Gold nanoparticle of diameter 30 nm (Sigma-Aldrich) was prepared by centrifuging the commercial aqueous nanoparticle solution in DI water solution at 13,000 rpm for 20 minutes to remove surfactants. Then, 10 μl of the clean nanoparticle solution with particle concentration 2E+11 /ml was drop cast on a clean glass surface. After waiting for 1 hour the sample was gently blow dried.

**References**

1. Nieto-Vesperinas, M. , Sáenz, J. J., Gómez-Medina, R. & Chantada, L. Optical forces on small magnetodielectric particles. *Opt. Express* **18**, 11428-11443 (2010).
2. Novotny, L. & Hecht, B. *Principles of Nano-Optics*. (Cambridge Univ. Press, Cambridge, UK, 2006).
3. Bohren, C. F., Huffman, D. R. Absorption and Scattering of Light by Small Particles. (*Wiley Interscience*, New York, 1983)
4. Rajapaksa, I. , Uenal, K. & Wickramasinghe, H. K. Image force microscopy of molecular resonance: A microscope principle. *Appl. Phys. Lett*. **97**, 073121 (2010).
5. Garcia, R. & Herruzo, E. T. The emergence of multifrequency force microscopy. *Nature Nanotech.* **7**, 217–226 (2012).